\documentclass[twocolumn,epjc3]{svjour3}

\usepackage[utf8]{inputenc}
\usepackage[T1]{fontenc}

\usepackage[numbers,sort&compress]{natbib}

\usepackage{microtype}

\usepackage{graphicx}
\usepackage{scalefnt}

\usepackage{amsmath}
\usepackage{amssymb}
\usepackage{amsfonts}
\usepackage{mathtools}
\usepackage{braket}

\usepackage{enumerate}
\usepackage{appendix}
\usepackage{empheq}
\usepackage{float}
\usepackage[parfill]{parskip}

\usepackage{lmodern}

\usepackage[pdftex,breaklinks]{hyperref}

\hypersetup{
    colorlinks = true,
    citecolor  = blue,
    linkcolor  = blue,
    urlcolor  = blue
}

\newcommand{\magicint}{``EM 1.8/2.0''}

\newcommand{\ai}{\textit{ab initio}}
\newcommand{\ie}{\textit{i.e.}}
\newcommand{\eg}{\textit{e.g.}}

\newcommand{\elem}[2]{$^{#2}$#1}

\newcommand{\ra}{\rangle}
\newcommand{\la}{\langle}

\begin{document}
\title{{\it Ab initio} Bogoliubov many-body perturbation theory}
\subtitle{Closed-form constraint on the average particle number}

\author{
P.~Demol\thanksref{ad:leuven,em:pd} \and 
T.~Duguet\thanksref{ad:irfu,ad:leuven,em:td} \and
A.~Tichai\thanksref{ad:tud,ad:emmi,ad:mpik,em:at}
}
\date{Received: \today{} / Accepted: date}

\thankstext{em:pd}{\email{pepijn.demol@kuleuven.be}}
\thankstext{em:td}{\email{thomas.duguet@cea.fr}}
\thankstext{em:at}{\email{alexander.tichai@physik.tu-darmstadt.de}}

\institute{%
\label{ad:leuven}%
KU Leuven, Instituut voor Kern- en Stralingsfysica, 3001 Leuven, Belgium
\and
\label{ad:irfu}%
IRFU, CEA, Universit\'e Paris-Saclay, 91191 Gif-sur-Yvette, France
\and
\label{ad:tud}%
Technische Universit\"at Darmstadt, Department of Physics, 64289 Darmstadt, Germany
\and 
\label{ad:emmi}%
ExtreMe Matter Institute EMMI, GSI Helmholtzzentrum f\"ur Schwerionenforschung GmbH, 64291 Darmstadt, Germany
\and
\label{ad:mpik}%
Max-Planck-Institut f\"ur Kernphysik, Saupfercheckweg 1, 69117 Heidelberg, Germany
}

\date{}

\allowdisplaybreaks

\maketitle

\begin{abstract}
Bogoliubov many-body perturbation theory (BMBPT) relying on the breaking of $U(1)$ global gauge symmetry has been recently formulated and applied to extend the applicability of standard perturbation theory to \ai{} calculations of atomic nuclei away from shell closures. So far, practical applications have been limited to second-order calculations due to the lack of a generic algorithm to constrain the average particle number of the symmetry-broken state. This limitation is presently lifted and a general BMBPT formalism is presented that allows to constrain the particle-number expectation value at arbitrary order $P$. The constraint can be incorporated in closed form by solving a polynomial equation of degree $P-1$. The numerical procedure is illustrated through BMBPT(3) calculations of calcium isotopes using a nuclear Hamiltonian derived within chiral effective field theory. 
\end{abstract}

\section{Introduction}
\label{intro}

The first-principles solution of the nuclear many-body problem has witnessed tremendous progress in the last years (see Refs.~\cite{Herg20review,Hebe203NF} for reviews). 
In particular the mild computational scaling of many-body expansion methods has allowed to extend the reach of state-of-the-art \ai{} approaches to heavy nuclei~\cite{Hu2022lead,Hebeler2023jacno,Tichai2023bcc,Arthuis2024}, collective features in open-shell systems~\cite{Frosini2021mrI,Frosini2021mrII,Hagen2022PCC,Porro:2024tzt,Giacalone:2024luz}, or electroweak processes~\cite{Miorelli2018,Kauf2068Ni,Yao18IMGCM}.
Within the portfolio of many-body formalisms, many-body perturbation theory (MBPT) provides a powerful method to describe finite nuclei (see Refs.~\cite{Tich16HFMBPT,Tichai18BMBPT,Tich19NatNCSM,Tichai2020review,Frosini2021mrIII}) and infinite nuclear matter~\cite{Hebe10nmatt,Dris17MCshort,Keller2021,Palaniappan2023}, offering an accurate yet computationally inexpensive way to compute (nuclear) observables.

The formulation of the Bogoliubov extension of MBPT (BMBPT) has enlarged the range of applicability to open-shell nuclei~\cite{Duguet2015u1,Arthuis2018adg1}. Today, realistic applications to both singly and doubly open-shell nuclei via spherical BMBPT~\cite{Tichai18BMBPT,Tichai2020review,Tichai2023bcc} and deformed BMBPT~\cite{Frosini2021,Scalesi:2024kms,Frosini:2024ajq} are routinely performed. In order to do so, BMBPT relies on the use of a Bogoliubov reference state breaking $U(1)$ symmetry associated with particle-number conservation. As a result, approximations to the exact ground-state of the system obtained by truncating the perturbative expansion are not eigenstates of the particle-number operator $A$ and the BMBPT expansion only recovers a good-symmetry state in the infinite-order limit. 
Hence, at any finite truncation order the particle number must be constrained to be correct {\it on average}, which can be done by driving the perturbative expansion with the grand potential  involving a Lagrange multiplier, the chemical potential, to be adjusted to satisfy the constraint~\cite{Duguet2015u1}. So far, the need to explicitly account for this constraint has been bypassed by restricting realistic applications to second-order BMBPT based on a canonical Hartree-Fock-Bogoliubov (HFB) reference state~\cite{RingSchuck}. Indeed, and as will be shown below, it is the only reference state and truncation order for which the perturbative correction to the reference state does not modify the average particle number and thus does not require a readjustment of the chemical potential beyond the mean field\footnote{The Bogoliubov reference state is itself adjusted to have the correct average particle number in the first place prior to correcting it in perturbation.}. 
\begin{figure}
    \includegraphics[width=0.95\columnwidth]{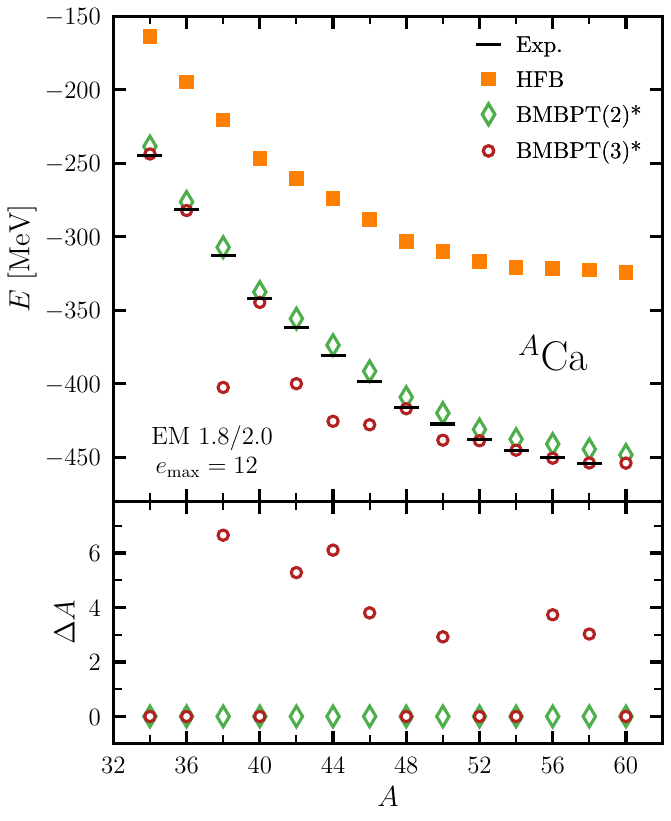}
    \caption{(Top panel) Total binding energy from HFB (orange squares), uncorrected BMBPT(2)* (green diamonds) and BMBPT(3)* (red circles) calculations are compared to experimental data. No adjustment of the average particle number is performed. (Bottom panel) BMBPT(2)* and BMBPT(3)* corrections to the average particle number. See Sec.~\ref{compdetails} for computational details.}
    \label{fig:ca-uncon}
\end{figure}

Figure~\ref{fig:ca-uncon} exemplifies the extent of the problem by displaying results of second- and third-order BMBPT calculations in Ca isotopes performed without particle number adjustment. As mentioned above, the correction to the average particle number is strictly zero in BMBPT(2) calculations and the systematic of binding energy along the isotopic chain is extremely reasonable and satisfying~\cite{Tichai18BMBPT,Tichai2023bcc}. Contrarily, the third-order correction shifts the average particle number by up to 6 neutrons in (neutron) open-shell isotopes. Such a particle number shift strongly reflects onto the binding energy of certain isotopes, making the corresponding BMBPT(3) masses highly unreasonable. It happens that the problem is enhanced by the fact that realistic nuclear interactions typically induce anomalously weak pairing at the mean-field level~\cite{Scalesi:2024kms,Frosini:2024ajq}, which itself yields a large enhancement of the particle-number corrections~\cite{Duguet2020zeropairing}.

In Ref.~\cite{Demol20BMBPT}, the consequence of the particle-number adjustment at each truncation order was systematically addressed in the context of schematic calculations. However, this work gave rise to two non-satisfactory solutions
\begin{itemize}
    \item[$i)$] a numerically costly iterative adjustment of the unperturbed reference state, 
    \item[$ii)$] an approximate {\it a posteriori} correction to the ground-state energy that cannot be propagated to other observables such as radii.  
\end{itemize} 

In this work, a fully satisfactory solution to this problem is provided and the general BMBPT formalism including the particle-number constraint is formulated at arbitrary orders. The constraint can be exactly incorporated into the formalism in closed algebraic form without invoking any costly iterative procedure. In practice, working out the constraint at order $P$ leads to solving a polynomial equation of degree $P-1$. The procedure is illustrated numerically by performing realistic \ai{} BMBPT(2) and BMBPT(3) (BMBPT(2,3)) calculations of calcium isotopes using two- and three-body interactions  derived within chiral effective field theory (EFT)~\cite{Epel09RMP,Mach11PR,Hamm13RMP,Hammer2020rmp}. The results are compared to those obtained from unconstrained BMBPT(2,3) calculations as well as from Bogoliubov coupled cluster (BCC) and valence-space in-medium renormalization group (VS-IMSRG) calculations.

On a longer term, going beyond such a scheme will require to explicitly restore $U(1)$ symmetry through the particle-number-projected BMBPT (PBMBPT) formalism~\cite{Duguet2015u1,Arthuis2020adg2} in order to obtain particle-number eigenstates at any truncation level. While this generalized formalism has not been implemented yet, numerical results of its coupled cluster counterpart can be found in Refs.~\cite{Qiu2019,Hagen2022PCC}. 

The present paper is organized as follows. The BMBPT formalism including the particle-number constraint is laid out in Sec.~\ref{sec:1} and specified to second- and third-order calculations. The results of \ai{} calculations are presented in Sec.~\ref{sec:results}. Conclusions are eventually provided in Sec.~\ref{sec:conclusions}.

\section{BMBPT formalism}
\label{sec:1}

\subsection{Bogoliubov algebra}

A $U(1)$-breaking Bogoliubov vacuum $|\Phi\ra$ is generated from the physical vacuum $|0\ra$ according to
\begin{align}
    |\Phi \ra = \mathcal{C} \prod_k \beta_k | 0 \ra \, ,
\end{align}
where $\mathcal{C}$ denotes a complex normalization constant.
The quasi-particle operators are obtained through a unitary Bogoliubov transformation~\cite{RingSchuck}
\begin{subequations}
\begin{align}
    \beta_k &= \sum_p U^\ast_{pk} c_p + V^\ast_{pk} c^\dagger_p \, ,\\
    \beta_k^\dagger &= \sum_p U_{pk} c^\dagger_p + V_{pk} c_p \, ,
\end{align}
\label{eq:bogotrafo}%
\end{subequations}
where the Fermionic quasi-particle operators fulfil the anti-commutation relations $\{\beta_p , \beta_q \} = \{ \beta^\dagger_p , \beta^\dagger_q\} = 0$ and $\{ \beta^\dagger_p , \beta_q \} = \delta_{pq} $. 

With the quasi-particle operators at hand, any, \eg{} two-body, operator $O$ such as the Hamiltonian\footnote{Even though the presently used Hamiltonian includes a three-nucleon interactions, it is effectively reduced to a two-body operator via  the rank-reduction method designed in Ref.~\cite{Frosini2021}.} of interest is normal ordered with respect to  $|\Phi\ra$ via the application of Wick's theorem~\cite{Wick50theorem} 
\begin{align}
    O & \equiv   O^{00}   \nonumber \\
    & \ + O^{20} + O^{11} + O^{02}  \nonumber \\
    & \ + O^{40} + O^{31} + O^{22} + O^{13} + O^{04} \ ,
\end{align}
where each contribution $O^{ij}$ involves $i$ ($j$) quasi-particle creation (annihilation) operators, \eg{},
\begin{align}
    O^{31} 
    = 
    \frac{1}{3!1!}
    \sum_{k_1k_2k_3k_4} o^{31}_{k_1k_2k_3k_4} \,
    \beta^\dagger_{k_1} \beta^\dagger_{k_2} \beta^\dagger_{k_3} \beta_{k_4} \, .
\end{align}
Explicit expressions of the matrix elements of each $O^{ij}$ in terms of the matrix elements of $O$ in the initial one-particle basis and of the Bogoliubov matrices $(U,V)$ can be found in Ref.~\cite{arthuis_bogoliubov_2018}. 

In practice, the Bogoliubov transformation and the associated $(U,V)$ matrices are obtained through a variational minimization in the manifold of Bogoliubov states under the constraint to have the correct neutron (proton) number in average, thus leading to the introduction of a neutron (proton) chemical potential via the grand potential
\begin{align}
    \Omega \equiv H - \lambda_N N - \lambda_Z Z \, ,
\end{align}
where $N$ ($Z$) denotes the neutron (proton) number operator. Doing so leads to solving the Hartree-Fock-Bogoliubov (HFB) mean-field equations and delivering the so-called canonical HFB state. A non-canonical HFB state can also be obtained by solving the HFB equations under additional constraints; \eg{} on the amount of pairing correlations carried by  $|\Phi\ra$~\cite{Duguet2020zeropairing}. In the following, neutron and proton numbers will not be explicitly distinguished such that $\lambda$ and $A$ refer to either one of them.

\subsection{Partitioning}

Being able to adjust the average particle number at each order $P$ of the BMBPT expansion\footnote{The counting of orders, i.e. the value of $P$, relates to the evaluation of the grand potential ground-state eigenvalue. Following this convention, the order $P\geq 1$ associated with a given observable corresponds to the order $(P-1)$ for the many-body state, \ie{} any observable counts for 1 order.} actually requires to make the chemical potential, and thus the grand potential, {\it order dependent}~\cite{Demol20BMBPT}, \ie{} for $P\geq 1$
\begin{equation}\label{eq:PNA-BMBPT_OmegaP}
\Omega^{(P)} \equiv H - \lambda^{(P)} A \, . 
\end{equation}
Given the unperturbed state $|\Phi \rangle $, the BMBPT expansion is based on a partitioning of the grand-potential operator\footnote{The operator $\Omega_0^{(P)}$ is the only one counting as a zeroth-order quantity.}
\begin{equation}
\Omega^{(P)} \equiv  \Omega_0^{(P)} + \Omega_{1}^{(P)} \, , \label{grandpot}
\end{equation}
such that $|\Phi \rangle $ and its elementary quasi-particle excitations are eigenstates of $\Omega_0^{(P)}$ for all $P$. As opposed to the iterative procedure put forward in Ref.~\cite{Demol20BMBPT}, the Bogoliubov unperturbed state remains {\it independent} of $P$ in the present scheme.
In this context, the unperturbed and residual parts of $\Omega^{(P)}$ read as
\begin{subequations}
    \begin{align}
    \Omega_0^{(P)} & \equiv \Omega^{00(P)} + \bar\Omega^{11} \\
    \Omega_1^{(P)} & \equiv \Omega^{20(P)} + \Breve\Omega^{11(P)} + \Omega^{02(P)}  \nonumber \\
    & \hspace{13pt} + \Omega^{40} + \Omega^{31} + \Omega^{22} + \Omega^{13} + \Omega^{04} \ , \label{eq:PNA-BMBPT-omega1P}
\end{align}
\end{subequations}
where the $P$ dependencies originating from the presence of $\lambda^{(P)}$ have been highlighted.  The $P$-independent diagonal one-body part of $\Omega_0^{(P)} $
\begin{equation}
    \bar\Omega^{11} \equiv \sum_k E_k \beta^\dagger_k \beta_k \, , \label{1BOmega0}
\end{equation} 
is built from a set of positive quasi-particle energies $E_k$ to be chosen.

Based on the above partitioning, standard algebraic or diagrammatic techniques can be employed to compute the BMBPT expansion of the ground-state wave function $| \Psi \ra$~\cite{Arthuis2018adg1}. Correspondingly, the ground-state energy or observable $O$ commuting with the Hamiltonian, \ie{} $[H, O] = 0$, can be expanded from a {\it projective} approach based on the asymmetric form $\mathcal{O} = \la \Phi | O | \Psi \ra$, where the ket (bra) is the fully correlated (unperturbed) state~\cite{Demol20BMBPT}.

\subsection{Particle-number constraint}

So far, the order-dependent chemical potential $\lambda^{(P)}$ has been left undetermined. Expanding the particle number, $\lambda^{(P)}$ is fixed by requiring that it matches the physical value A at order $P$, \ie{},
\begin{equation} \label{eq:PNA-BMBPT_A}
    A^{[P]} \equiv \langle \Phi | A  \sum_{p=0}^{P-1} \left [ R_0\,
    \Omega_{1}^{(P)} \right ]^{p} | \Phi \rangle_{\text{C}} = \text A \ ,
\end{equation}
where the subscript `C' specifies that only {\it connected} terms or diagrams must be retained~\cite{Duguet2015u1,Arthuis2018adg1} and where the resolvent operator
\begin{equation}
    R_0 \equiv \Big[\Omega^{00(P)} - \Omega_0^{(P)} \Big]^{-1} = -(\bar\Omega^{11})^{-1} \, ,
\end{equation}
happens to be independent of $\lambda^{(P)}$. Since $\Omega_{1}^{(P)}$ depends linearly on $\lambda^{(P)}$, Eq.~\eqref{eq:PNA-BMBPT_A} is an implicit polynomial equation of degree $P-1$ in $\lambda^{(P)}$. 

The reference state $|\Phi \rangle $ having been constrained to carry the physical particle number on average,  \ie{} $A^{[1]}=\langle \Phi | A | \Phi \rangle = \text A$, Eq.~\eqref{eq:PNA-BMBPT_A} actually demands that the {\it shift} of the average particle number at order $P$ is zero, \ie{}  
\begin{equation} 
\Delta A^{[P]} \equiv A^{[P]} - A^{[1]} = 0 \, .
\end{equation} 
Thus, the scheme is more conveniently re-expressed via the identity 
\begin{equation}   
\Omega^{(P)} = \Omega^{(1)} - \Delta \lambda^{(P)} A \, , \label{rewriteOmegaP}
\end{equation}
making use of the $P$th-order chemical potential {\it shift}
\begin{equation}   
\Delta \lambda^{(P)} \equiv \lambda^{(P)} - \lambda^{(1)} \, , 
\end{equation}   
such that
\begin{subequations}
\label{eq:BMBPT-omegaP}
\begin{align}
\Omega_0^{(P)} 
    & \equiv \Omega_0^{(1)} - \Delta \lambda^{(P)} A^{00},  \\
    \Omega_1^{(P)} \label{eq:BMBPT-omega1P}
    & \equiv  \Omega_1^{(1)}  - \Delta \lambda^{(P)} A_N  ,
\end{align}
\end{subequations}
where $A_N \equiv A -  A^{00}$. 
Exploiting Eqs.~\eqref{rewriteOmegaP}-\eqref{eq:BMBPT-omegaP}, the particle-number shift can be rewritten as
\begin{align}
    \Delta A^{[P]} &= \langle \Phi | A_N \sum_{p=1}^{P-1} \left [ R_0 \,
    (\Omega_1^{(1)} - \Delta \lambda^{(P)} A_N) \right ]^{p} | \Phi \rangle_{\text{c}}  \nonumber \\
    &= 0 \label{finalformalexpression}
\end{align}
making explicit that it yields in fact a polynomial equation of degree $P-1$ in $\Delta \lambda^{(P)}$.

Once $\lambda^{(P)}$ has been obtained by solving Eq.~\eqref{finalformalexpression}, the $P$th-order grand potential $\Omega^{(P)}$ is effectively defined through Eq.~\eqref{eq:PNA-BMBPT_OmegaP}. While the average particle number $A^{[P]}$ is equal to A by construction, the energy is eventually obtained from
\begin{align}
    E^{[P]}  &= \langle \Phi | H \sum_{p=0}^{P-1} \left ( R_0 \, \Omega_{1}^{(P)} \right )^{p} | \Phi \rangle_{\text{c}} \notag \\
     &= \langle \Phi | H | \Phi \rangle + \langle \Phi | \Omega_{1}^{(P)} \sum_{p=1}^{P-1} \left ( R_0 \, \Omega_{1}^{(P)} \right )^{p} | \Phi \rangle_{\text{c}}  \, ,
     \label{eq:PNA-BMBPT_E}
\end{align}
where the perturbative corrections can be expressed in terms of the sole residual grand potential $\Omega_{1}^{(P)}$.

\subsection{Low-orders algebraic expressions}

\begingroup
\tolerance=10000
Using established diagrammatic techniques for BMBPT~\cite{Duguet2015u1,Arthuis2018adg1}, explicit algebraic expressions for the low-order particle-number shift (Eq.~\eqref{finalformalexpression}) are obtained in terms of second-quantized matrix elements of the quasi-particle operators.
To this end, one needs to evaluate all possible BMBPT diagrams of order $p+1$
\begin{align*}
    \text{Diagram}^{(p+1)}_i (A_N, \underbrace{\Omega_{1}^{(1)}}_{m}, \underbrace{A_N}_{p-m})
\end{align*}
with $1\leq p < P$, containing $p+1$ vertices among which $A_N$ is the bottom vertex, $m$ vertices originate from $\Omega_{1}^{(1)}$ whereas the  $(p-m)$ remaining vertices correspond to $A_N$. This set is obtained from the BMBPT diagrams contributing to the average particle number calculated with the {\it first-order} grand potential~\cite{Arthuis2018adg1} by replacing {\it one-body} vertices originating from $\Omega_{1}^{(1)}$ with their $A_N$ counterparts in all possible ways.
The computational scaling of the diagram evaluation is subdominant compared to the energy since $A_N$ is only a one-body operator. Hence the calculation of the chemical potential shift involves only a small fraction of the total runtime and is much lower than for the method based on the iterative adjustment of the reference state introduced in Ref.~\cite{Demol20BMBPT}.
\endgroup

\subsection{BMBPT(2,3) shifted chemical potential}

In BMBPT(2) calculations, the particle-number shift is obtained from the roots of a polynomial of degree one and hence requires only the trivial solution of a linear equation.
Since BMBPT diagrams evaluate to real numbers, the corresponding solution for the chemical potential shift is a real number itself and so is the grand potential
\begin{align}
    \Omega_{1}^{(2)} 
    = \Omega_1^{(1)} - \Delta \lambda^{(2)} A \, .
\end{align}
As proven in App.~\ref{app:pna3}, the particle-number shift at second order is zero, \ie{} $\Omega_{1}^{(2)} = \Omega_{1}^{(1)}$, when using a canonical reference state.

In BMBPT(3) calculations, the chemical potential shift emerges as the root of a polynomial of degree two
\begin{align}
    C_2 \left(\Delta \lambda^{(3)}\right)^2 +
    C_1 \Delta \lambda^{(3)} +
    C_0
    =
    0 \, ,
    \label{eq:quad_C}
\end{align}
where the explicit expressions of the coefficients $C_0, C_1$ and $C_2$ can be found in App.~\ref{app:pna3}.
Since quadratic equations yield between zero and two real solutions, the identification of the physical solution for $\Delta \lambda^{(3)}$ is unclear. While the presence of zero real solution signals an ill-defined BMBPT expansion, two real solutions make necessary to pin down the `physical' one. In cases where the quadratic term $C_2$ is small, one may anticipate the correct solution to be close to the one obtained by solving the linear approximation to the quadratic equation.

\section{Numerical results}
\label{sec:results}

\subsection{Computational details}
\label{compdetails}

All numerical simulations are performed employing a spherical harmonic oscillator (HO) one-body basis characterized by the oscillator frequency $\hbar \omega=16\mathrm{\,MeV}$. The single-particle basis is truncated according to $e_{\mathrm{max}} = (2n + l)_\text{max}=12$, where $n$ denotes the radial quantum number and $l$ the orbital angular momentum, respectively. 

The \magicint{} Hamiltonian presently employed includes a chiral N${}^{3}$LO nucleon-nucleon (2N) interaction evolved to low-momentum scale \(\lambda_{\text{SRG}}=1.8\text{ fm}^{-1}\) via a similarity renormalization group transformation~\cite{Bogn05nuclmat,Jurg09SRG3N}. The NN force is supplemented by a N${}^{2}$LO three-nucleon (3N) interaction with cutoff $\Lambda=2.0\,\mathrm{fm}^{-1}$ whose low-energy constants are adjusted to $A=3,4$ observables, as described in Ref.~\cite{Hebe11fits}. The three-body interaction is represented in the three-body HO basis truncated at $E_{3\mathrm{max}}= 24$, which was shown to be sufficient to obtain convergence in nuclei up to mass number $A\lesssim 100$~\cite{Miyagi2021}. Chiral matrix elements were obtained using the \texttt{Nuhamil} code~\cite{Miyagi2023nuhamil}. As mentioned earlier, the three-body Hamiltonian is approximated as an effective two-body operator by applying the rank-reduction procedure described in Ref.~\cite{Frosini2021}.

The many-body vacuum $| \Phi \rangle$ is obtained from a spherical HFB calculation and hence is an eigenstate of $\mathbf{J}^2$ and $\mathbf{J}_z$ operators.

\subsection{Particle-number constraint at third order}

The chemical-potential shift obtained in BMBPT(3) calculations based on a canonical HFB state is now characterized. The weak pairing correlations induced by chiral interactions generate small quasi-particle energies and can thus induce large particle-number corrections as was already seen in Fig.~\ref{fig:ca-uncon}. 

In this context, Eq.~\eqref{eq:quad_C} may not always admit a real solution when using the HFB quasi-particle energies, as traditionally done, to define the one-body part $\bar{\Omega}^{11}$ of the unperturbed grand potential $\Omega_0^{(3)}$. To anticipate such a problem, the energies entering  $\bar{\Omega}^{11}$ in Eq.~\eqref{1BOmega0} are presently chosen according to
\begin{align}
    E_k= \Omega^{11(1)}_{kk} + \Delta_E \, , \label{shiftdef}
\end{align}
\ie{} an arbitrary global shift $\Delta_E$ is added to the diagonal part of $\Omega^{11(1)}$\footnote{When using a canonical HFB state as done in the present section, $\Omega^{11(1)}$ is diagonal with eigenvalues corresponding to the set of HFB quasi-particle energies.}. 

\begin{figure}[t]
    \centering
    \includegraphics[width=0.85\columnwidth]{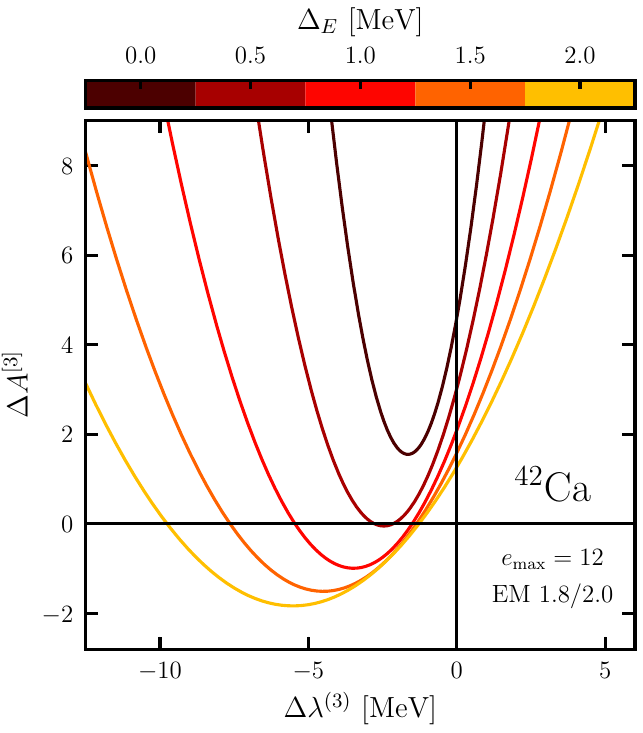}
    \caption{Second-order polynomial delivering the particle-number shift in BMBPT(3) calculations of \elem{Ca}{42} as a function of $\Delta \lambda^{(3)}$ for five values of the shift $\Delta_E$ in Eq.~\eqref{shiftdef}.}
    \label{fig:roots}
\end{figure}

Varying $\Delta_E$ provides a diagnostic tool to characterize the capacity to constrain BMBPT(3) calculations to display the correct particle number on average depending on the employed  partitioning of the grand potential. This is illustrated in Fig.~\ref{fig:roots} for $^{42}$Ca where the second-order polynomial on the left-hand side of Eq.~\eqref{eq:quad_C}, \ie{} the particle-number shift, is displayed as a function of $\Delta \lambda^{(3)}$ for five values of $\Delta_E$. Using the traditional partitioning, \ie{} $\Delta_E=0$, the quadratic function does not cross the horizontal axis such that Eq.~\eqref{eq:quad_C} does not display any solution. As such, $^{42}$Ca is a rare ``pathological'' case along the Ca isotopic chain\footnote{Between $^{34}$Ca and $^{60}$Ca, the only two isotopes for which no solution to the quadratic equation exists are $^{42}$Ca and $^{56}$Ca.} when using the traditional partitioning. Increasing $\Delta_E$, the quadratic function is lowered and crosses the horizontal axis for $\Delta_E=0.5$\,MeV such that a meaningful BMBPT(3) calculation of $^{42}$Ca can indeed be performed for $\Delta_E \geq 0.5$\,MeV. One observes that the root closest to zero remains very stable as $\Delta_E$ increases, whereas the other solution decreases monotonically. 

\begin{figure}[t]
    \centering
    \includegraphics[width=0.85\columnwidth]{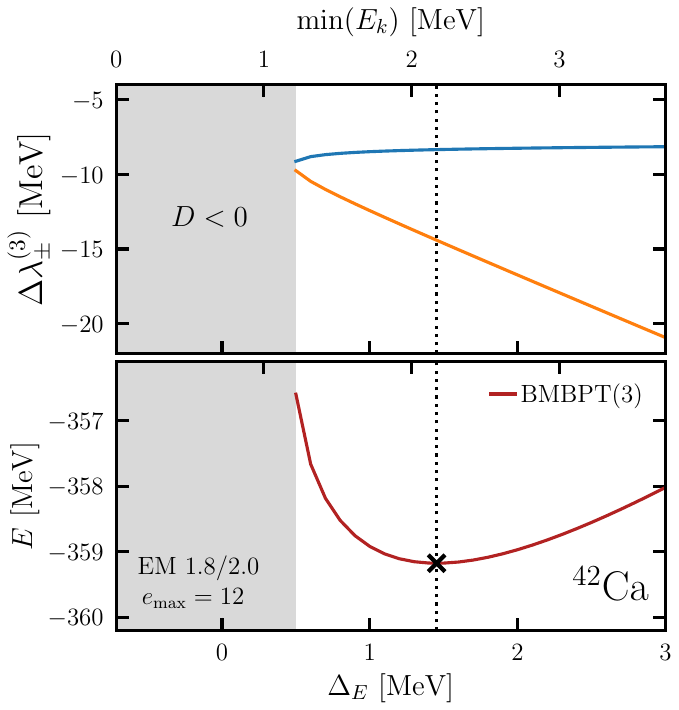}
    \caption{Roots of the third-order chemical potential shift (top) and ground-state energy (bottom) as a function of the denominator shift for \elem{Ca}{42}. The gray area indicates a domain where no real root was obtained.}
    \label{fig:stationarity}
\end{figure}

The above features can be clearly seen from the upper panel of Fig.~\ref{fig:stationarity}. The fact that this solution of the quadratic equation corresponds to the smallest chemical potential shift and remains largely insensitive to the arbitrary energy displacement  $\Delta_E$ makes it the physically-preferred solution for $\Delta \lambda^{(3)}$.

With this solution at hand for $\Delta \lambda^{(3)}$, it is still unclear which value of $\Delta_E$ should be preferred. One can make an educated choice by looking at the lower panel of Fig.~\ref{fig:stationarity} showing the BMBPT(3) binding energy of  \elem{Ca}{42} as a function of $\Delta_E$. It is observed that increasing $\Delta_E$ beyond $0.5$, the energy displays a minimum
\begin{align}
    \frac{\partial E}{\partial \Delta_E} = 0 \, ,
\end{align}
for $\Delta_E \approx 1.4$\,MeV around which the energy is thus the least sensitive to the value taken by that parameter. In the following such a stationarity criterion is employed in all calculations, including nuclei for which a solution could be found for $\Delta_E = 0$. Because results will only be shown for such a choice of $\Delta_E$, it is worth mentioning that this choice indeed delivers the best phenomenological BMBPT(3) results.

\begin{figure}[t!]
    \centering
    \includegraphics[width=0.95\columnwidth]{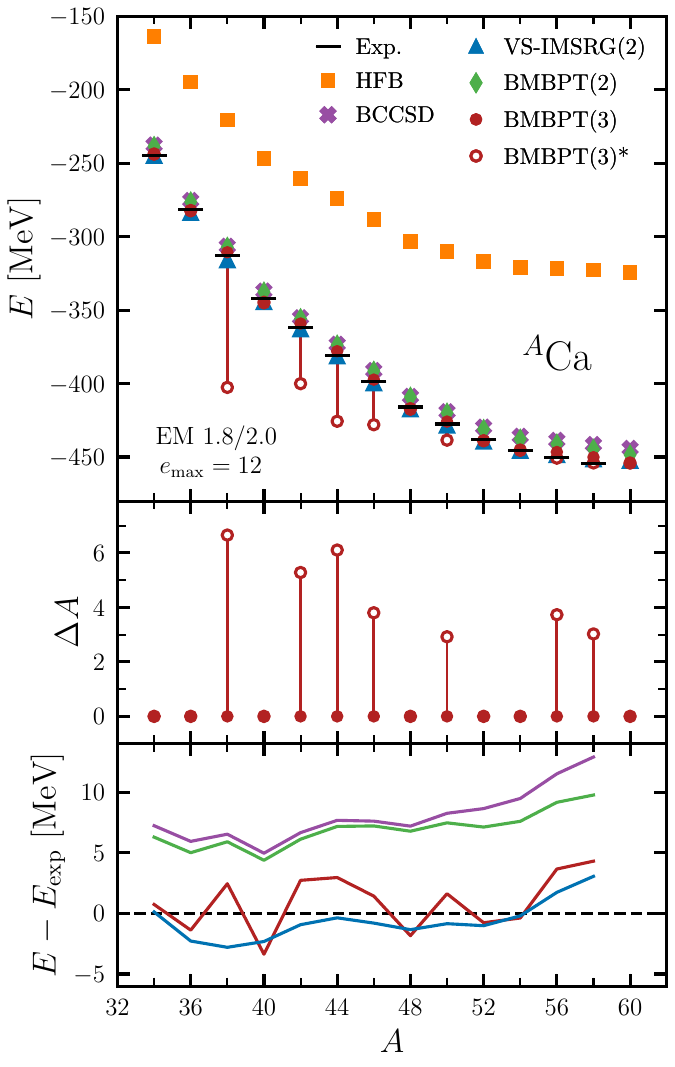}
    \caption{Ground-state energy (top), particle-number correction (middle) and energy difference to experimental data (bottom) for HFB, BMBPT(2,3), BMBPT(3)*, BCCSD and VS-IMSRG(2) calculations along the Ca isotopic chain.}
    \label{fig:caEgs}
\end{figure}

\subsection{BMBPT(2,3) calculations of Ca isotopes}

Now that BMBPT(2,3) calculations based on both canonical and non-canonical HFB state can be meaningfully performed, simulations based on the former reference state are compared along the calcium isotopic chain to state-of-the-art non-perturbative many-body approaches, \ie{}, Bogoliubov coupled-cluster theory~\cite{Tichai2023bcc} and the valence-space in-medium similarity renormalization group (VS-IMSRG) approach~\cite{Bogn14SM,Stroberg2019}.
Relying on a full diagonalization in the active-space, the VS-IMSRG is presently considered to be the most accurate method of all. 

Figure~\ref{fig:caEgs} displays the ground-state energy (top), particle-number correction (middle) and energy difference to experimental data (bottom) for BMBPT(2,3), BMBPT(3)*, BCCSD and VS-IMSRG(2) calculations for $^{34-60}$Ca. As already discussed in connection with Fig.~\ref{fig:ca-uncon}, unconstrained BMBPT(3)* calculations are severely contaminated by the lack of particle-number adjustment. In particular, the binding energies of \elem{Ca}{38} and \elem{Ca}{42,44,46} yield a significant overbinding due to the large particle-number shift of $\Delta A \approx 4-6$ particles that translates into up to $70\,$MeV of extra binding. This is also observed to a lesser extent in \elem{Ca}{50}.

By adjusting the third-order chemical potential, the pathologies characterizing BMBPT(3)* results are resolved such that BMBPT(3) results quantitatively follow VS-IMSRG(2) and BCCSD trends. For all nuclei BMBPT(3) simulations give a sizeable attractive correction to BMBPT(2) of the order of $10\%$ of the BMBPT(2) correlation energy.
The BMBPT(3) results are in closer agreement with VS-IMSRG(2) than BCCSD. This is attributed to an overprediction of attractive two-particle-two-hole correlations at third order that is actually corrected for by the resummation operated at the BCCSD level.

\begin{figure}[t]
    \centering
    \includegraphics[width=0.95\columnwidth]{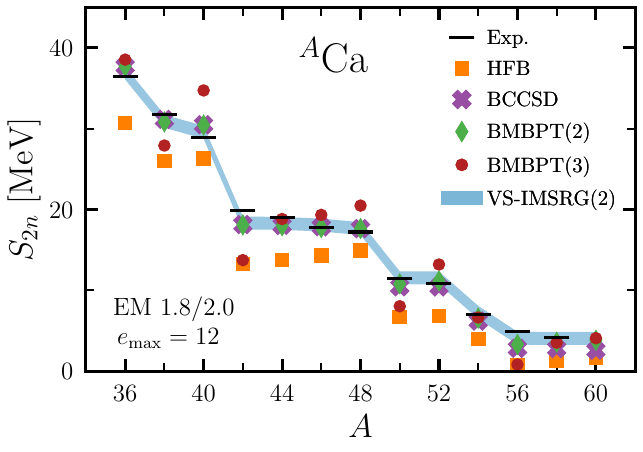}
    \caption{Two-neutron separation energies along calcium isotopes.}
    \label{fig:caS2n}
\end{figure}

The accurate low-order BMBPT predictions of bulk properties directly reflect the softness of the  Hamiltonian~\cite{Tich16HFMBPT,Tichai2020review}. The  \magicint{} Hamiltonian is indeed known to be amenable to perturbative many-body simulations and this feature clearly extends to open-shell nuclei~\cite{Tichai2023bcc}.
The softness of the interaction is already visible at the mean-field level where the HFB reference states capture the bulk of the total binding energy (60-70\%) throughout the entire calcium chain.

While low-order (unprojected) BMBPT calculations offers an inexpensive way to extract bulk properties of atomic nuclei, the corresponding results for differential quantities may become  sensitive to the choice of the reference state and the partitioning of the grand potential.
To this end, two-neutron separation energies
\begin{align}
    S_{2n} (N,Z) = E(N,Z) - E(N-2,Z) \, ,
\end{align}
obtained for Ca isotopes from the various many-body methods are displayed in Fig.~\ref{fig:caS2n}. For all isotopes  up to the most neutron-rich nucleus \elem{Ca}{58} measured experimentally, BMBPT(2) and the more sophisticated BCCSD or VS-IMSRG(2) predictions  nicely follow the experimental trend\footnote{For a more detailed analysis of such results and of the associated phenomenology, the interested reader is referred to Ref.~\cite{Scalesi:2024kms}.}. 

Including the third-order correction, the behavior of the $S_{2n}$ deteriorates significantly in the vicinity of magic numbers, \eg{} $N=20,28$, whereas in between them, \eg{} for \elem{Ca}{44,46}, BMBPT(3) values are in good agreement with BCCSD/VS-IMSRG(2) predictions and experimental data. Eventually, the perturbative expansion is too sensitive to the particle-number adjustment to compute refined differential energies involving nuclei for which the extent of the particle-number breaking differ notably, \ie{} near shell closures. Contrarily, one observes that BCC results, which also rely on an order-dependent adjustment of the average particle number~\cite{demol25}, are much more robust thanks to the non-perturbative character of the method and to the self-consistent determination of the cluster amplitudes that more gently absorb the impact associated with the chemical potential shift.

\section{Conclusions}
\label{sec:conclusions}

The present work generalizes Bogoliubov many-body perturbation theory to account, at any truncation order, for the correct average particle number. This necessity is due to the breaking of $U(1)$ symmetry that characterizes (truncated) BMBPT. Practically, the constraint is satisfied at order $P$ by solving a polynomial equation of degree $P-1$ to determine the value of the appropriate chemical potential.

While BMBPT(2) calculations published so far did not require any particle number adjustment and delivered very satisfactory results, the first consistent BMBPT(3) results are presently performed along the Ca isotopic chain based on chiral two- and three-nucleon interactions. Because of the weak pairing correlations induced by realistic nuclear interactions at the mean-field level, the shift in particle number to be compensated for can be anomalously large in certain open-shell isotopes such that no solution for the necessary chemical potential shift exists in a few of them. The difficulty is systematically overcome by allowing for a more general partitioning than before when setting up the perturbative expansion\footnote{The situation can also be improved by constraining the HFB reference state to carry stronger pairing correlations via the use of a dedicated Lagrange term~\cite{Duguet2020zeropairing}. In such a case, the unperturbed state is non-canonical, \ie{}, $\Omega^{20}$, $\Omega^{02}$ and ${\breve \Omega}^{11}$ are different from zero. While doing so helps, it does not offer a systematic and black-box resolution of the issue, which is why a different option is put forward in the present work.}.

Based on the above algorithm, BMBPT(3) binding energies of Ca isotopes were shown to be on par with those obtained from sophisticated Bogoliubov coupled cluster and in-medium similarity renormalization group calculations. The accuracy on more refined differential quantities such as two-neutron separation energies is however significantly lower for BMBPT(3). This shortcoming is due to the anomalous sensitivity of the perturbative results to the particle-number adjustment in the vicinity of shell closures. Contrarily, non-perturbative Bogoliubov coupled cluster calculations, which also require an order-dependent adjustment of the average particle number~\cite{demol25}, are seen to be much more robust in this respect. As for perturbation theory, going beyond BMBPT(2) in a fully satisfactory fashion thus calls for an explicit restoration of $U(1)$ symmetry via projected BMBPT~\cite{Duguet2015u1}. In the meantime, all-order methods such as \eg{} Bogoliubov coupled cluster theory provide a numerically more robust and diagrammatically more complete account of correlations and hence will be favoured in applications dedicated to high-accuracy simulations.

\begin{acknowledgements}
We thank T. Miyagi for providing us with matrix elements for the employed chiral interactions.
This work was supported in part by Research Foundation Flanders (FWO, Belgium, grant 11G5123N).
The work of A.T.~was supported by the European Research Council (ERC) under the European Union's Horizon 2020 research and innovation programme (Grant Agreement No.~101020842).
Computations were in part performed with an allocation of computing resources at the J\"ulich Supercomputing Center. 
Additional computational resources and services used in this work were provided by the VSC (Flemish Supercomputer Center), funded by the Research Foundation - Flanders (FWO) and the Flemish Government.
\end{acknowledgements}

\begin{appendices}

\section{Algebraic expressions}
\label{app:pna3}

In this appendix, explicit algebraic expressions of the coefficients entering the polynomial (Eq.~\eqref{finalformalexpression}) to be solved in order to obtain $\Delta \lambda^{(P)}$ are provided for BMBPT(2,3). 

As discussed in the main body of the text, the second-order polynomial equation to be solved in BMBPT(3) calculations 
\begin{align}
    C_2 \left(\Delta\lambda^{(3)}\right)^2
    + C_1 \Delta\lambda^{(3)}
    + C_0 = 0 \, ,
\end{align}
is expressed in terms of coefficients that can be evaluated diagrammatically. Below, the short-hand notation
\begin{subequations}
\begin{align}
    E_{k_1k_2} &\equiv E_{k_1} + E_{k_2} \, , \\
    E_{k_1k_2k_3k_4} &\equiv E_{k_1} + E_{k_2} +  E_{k_3} + 
    E_{k_4} \, ,
\end{align}
\end{subequations}
for two- and four-quasi-particle energy denominators is employed.

The coefficient $C_2$ of the quadratic term is given by
\begin{align}
    C_2 
    \equiv
    \sum_{k_1k_2k_3} \dfrac{A^{20}_{k_1k_2}A^{11}_{k_3k_1}A^{02}_{k_3k_2}}{E_{k_1 k_2}E_{k_2k_3}} \, .
\end{align}

Next, the coefficient $C_1$ of the linear term sums the five contributions
\begin{subequations}
\begin{align}
C^{[1]}_1
&= 
\dfrac{1}{2} \sum_{k_1k_2}\dfrac{A^{20}_{k_1k_2}A^{02}_{k_1k_2}}{E_{k_1k_2}}  \, , \\
C^{[2]}_1 
&=   
-\sum_{k_1k_2k_3} \dfrac{A^{20}_{k_1k_2}\breve\Omega^{11(1)}_{k_3k_1}A^{02}_{k_3k_2}}{E_{k_1k_2}E_{k_2k_3}} \, , \\
C^{[3]}_1
&=
-\sum_{k_1k_2k_3} \dfrac{A^{20}_{k_1k_2}A^{11}_{k_3k_1}\Omega^{02(1)}_{k_3k_2}}{E_{k_1k_2}E_{k_2k_3}} \, , \\
C^{[4]}_1
&=
-\dfrac{1}{4}\sum_{k_1k_2k_3k_4}\dfrac{A^{20}_{k_1k_2}\Omega^{22}_{k_3k_4k_1k_2}A^{02}_{k_3k_4}}{E_{k_1k_2}E_{k_3k_4}} \, , \\
C^{[5]}_1     
&=
-\dfrac{1}{4}\sum_{k_1k_2k_3k_4}\dfrac{A^{20}_{k_1k_2}\Omega^{04}_{k_1k_2k_3k_4} A^{20}_{k_3k_4}}{E_{k_1k_2}E_{k_1k_2k_3k_4}} \, .
\end{align}
\end{subequations}
Similarly the constant term $C_0$ sums the five contributions
\begin{subequations}
\begin{align}
C^{[1]}_0
&=
-\dfrac{1}{2} \sum_{k_1k_2}\dfrac{A^{20}_{k_1k_2}\Omega^{02(1)}_{k_1k_2}}{E_{k_1k_2}} \, , \\
C^{[2]}_0
&=
\sum_{k_1k_2k_3} \dfrac{A^{20}_{k_1k_2}\breve\Omega^{11(1)}_{k_3k_1}\Omega^{02(1)}_{k_3k_2}}{E_{k_1k_2}E_{k_2k_3}} \, , \\
C^{[3]}_0
&=     
\dfrac{1}{4}\sum_{k_1k_2k_3k_4}\dfrac{A^{20}_{k_1k_2}\Omega^{22}_{k_3k_4k_1k_2}\Omega^{02(1)}_{k_3k_4}}{E_{k_1k_2}E_{k_3k_4}} \, , \\
C^{[4]}_0
&=
\dfrac{1}{4}\sum_{k_1k_2k_3k_4}\dfrac{A^{20}_{k_1k_2}\Omega^{04}_{k_1k_2k_3k_4}\Omega^{20(1)}_{k_3k_4}}{E_{k_1k_2} E_{k_1k_2k_3k_4}} \, , \\
C^{[5]}_0
&=
\dfrac{1}{6}\sum_{k_1k_2k_3k_4k_5} \dfrac{A^{20}_{k_1k_2}\Omega^{31}_{k_3k_4k_5k_1}\Omega^{04}_{k_3k_4k_5k_2}}{E_{k_1k_2} E_{k_2k_3k_4k_5}} \, .
\end{align}
\end{subequations}

The linear equation at play in BMBPT(2) calculations is obtained from the above by realizing that only $C^{[1]}_1$ and $C^{[1]}_0$ contribute in this case. Because $\Omega^{02(1)}=0$ whenever using a canonical HFB state~\cite{Duguet2015u1,Arthuis2018adg1}, $C^{[1]}_0$ vanishes and $\Delta\lambda^{(2)}=0$ in this particular setting.

\end{appendices}

\bibliographystyle{apsrev4-1}
\begingroup
\small
\newcommand{\urlprefix}{}
\bibliography{references}
\endgroup

\end{document}